# Monolithic lithium niobate photonic chip for efficient terahertz-optic modulation and terahertz generation

Yiwen Zhang[1,2,†], Jingwei Yang[1,2,3,†,*], Zhaoxi Chen[1,2], Hanke Feng[1,2], Sha Zhu[4], Kam-Man Shum[2], Chi Hou Chan[1,2] & Cheng Wang[1,2,3,*]

1 Department of Electrical Engineering, City University of Hong Kong, Kowloon, Hong Kong, China
2 State Key Laboratory of Terahertz and Millimeter Waves, City University of Hong Kong, Kowloon, Hong Kong, China
3 Hong Kong Institute for Advanced Study, City University of Hong Kong, Kowloon, Hong Kong, China
4 Institute of Intelligent Photonics, Nankai University, Tianjin, 300071, China
†These authors contributed equally to this work.
*Corresponding authors: jingwyang7@cityu.edu.hk; cwang257@cityu.edu.hk

**Abstract**
The terahertz (THz) frequency range, bridging the gap between microwave and infrared frequencies, presents unparalleled opportunities for advanced imaging, sensing, communications, and spectroscopy applications. Terahertz photonics, in analogy with microwave photonics, is a promising solution to address the critical challenges in THz technologies through optical methods. Despite its vast potential, key technical challenges remain in effectively interfacing THz signals with the optical domain, especially THz-optic modulation and optical generation of THz waves. Here, we address these challenges using a monolithic integrated photonic chip designed to support efficient bidirectional interaction between THz and optical waves. Leveraging the significant second-order optical nonlinearity and strong optical and THz confinement in a thin-film lithium niobate on quartz platform, the chip supports both efficient THz-optic modulation and continuous THz wave generation at up to 500 GHz. The THz-optic modulator features a radio frequency (RF) half-wave voltage of 8V at 500 GHz, representing more than an order of magnitude reduction in modulation power consumption from previous works. The measured continuous wave THz generation efficiency of $4.8 \times 10^{-6}$ /W at 500 GHz also marks a tenfold improvement over existing tunable THz generation devices based on lithium niobate. We further leverage the coherent nature of the optical THz generation process and mature optical modulation techniques to realize high-speed electro-THz modulation at frequencies up to 35 GHz. The chip-scale THz-photonic platform paves the way for more compact, efficient, and cost-effective THz systems with potential applications in THz communications, remote sensing, and spectroscopy.

**Introduction**
Terahertz (THz) wave, spanning 0.1-10 THz in the electromagnetic spectrum, is increasingly recognized for its potential applications across diverse fields, including high-speed wireless (6G) communications[1], medical imaging[2], chemical identification[3], THz radar[4], and THz time-domain spectroscopy[5]. However, current THz technology still faces critical challenges including the lack of efficient THz emitters and detectors, excessive loss in cables and electronic devices at high frequencies, and difficulties in achieving more compact and integrated THz devices and systems[6]. THz photonics, drawing inspiration from microwave photonics, could potentially provide a compelling solution to these challenges by generating, transmitting and processing THz signals in the optical domain[7]. This is because photonic systems typically feature low transmission loss, no inherent gain-bandwidth limitations, and convenience for large-scale integration[8]. Central to such THz photonic systems are THz-optic modulators that coherently convert THz signals into the optical domain and optical generation of THz waves for conversion back to THz frequencies. Importantly, such coherent and bidirectional conversion processes allow the use of low-cost and versatile optical techniques and instruments like lasers, modulators and fibers for the currently challenging tasks of control, manipulation, and transmission of THz signals.

On the THz-optic modulation front, traditional electro-optic modulator platforms, such as silicon (Si)[9], indium phosphide (InP)[10], and bulk lithium niobate (LiNbO$_3$, LN), face significant challenges in achieving modulation frequencies beyond 100 GHz. Recently, modulators based on plasmonic structures have reached modulation bandwidths of 500 GHz[11]. However, these plasmonic modulators usually exhibit significant optical losses of 0.4-0.5 dB/μm and limited power handling capabilities[11,12]. Intrinsically, LN is an excellent material for THz-optic modulation due to its large and almost instantaneous electro-optic response ($r_{33}$= 30.8 pm/V), high optical damage threshold, and low optical losses[13]. However, traditional bulk LN modulators are typically limited to bandwidths around 35 GHz due to their long device lengths, leading to exacerbated RF loss at increased frequencies[14]. The thin-film LN (TFLN) platform has recently emerged as a promising candidate to achieve ultrabroad-band modulators thanks to the dramatically improved electro-optic efficiencies in high-confinement waveguides and much reduced RF loss. Further combining electro-optic modulators with other high-performance functional devices on the same TFLN platform, including high-quality ($Q$) resonators[15,16] and frequency combs[17] as well as low-loss delay lines[18], has led to the realization of ultrahigh-speed and low-power microwave photonics[19]. Such TFLN microwave photonic systems could be readily extended to the THz range if efficient THz-optic modulation could be achieved. Currently, many state-of-the-art TFLN modulators can support 3 dB bandwidths beyond 100 GHz[20-27], in particular those with advanced capacitively loaded electrodes[25,26] for more flexible control of RF velocity and loss. Yet very few of them have experimentally validated performances deep into the THz bands due to difficulties associated with measurements at ultrahigh frequencies[22]. Our previous work has demonstrated an RF $V_\pi$ of 7.3V at 250 GHz, which however deteriorates quickly beyond this frequency and would have extrapolated to around 28 V at 500 GHz, mainly due to substantial transmission line losses and residual velocity mismatch at ultrahigh frequencies[28]. In short, challenges persist in achieving efficient modulation and low RF $V_\pi$ in the THz band, especially above 300 GHz.

On the THz generation front, various optical methods have been proposed and demonstrated. The difference frequency generation (DFG) method, which involves mixing two optical signals separated by the desired THz frequency via a second-order ($\chi^{(2)}$) nonlinearity, is particularly interesting for THz photonics since it allows a seamless and coherent conversion between optics and THz. Among various $\chi^{(2)}$ nonlinear material platforms, such as LN[29-36], gallium phosphide (GaP)[37], gallium arsenide (GaAs)[38], zinc telluride (ZnTe)[39], gallium selenide (GaSe)[40], and zinc germanium phosphide (ZGP)[41], LN stands out with its large $\chi^{(2)}$ coefficients, relatively high optical damage threshold, and ultra-low optical absorption in the near-infrared spectral range. Most of these devices make use of pulsed optical input to increase the instantaneous pump power and produce pulsed THz radiation[30-33], which is ideal for ultrafast imaging and time-domain spectroscopy. However, many other applications such as 6G wireless communication and biomedical imaging, necessitate continuous THz wave sources for uninterrupted radiation and non-destructive testing[42]. Although there have been successful attempts at continuous THz wave generation[34-36], these methods often require high optical input power of tens of watts or more due to the weak nonlinear interactions in bulk crystals. This leads to considerable power consumption and relatively low conversion efficiency (approximately $10^{-7}$/W[35,36]). Additionally, these devices feature large footprints and require complex and extensive experimental conditions, limiting their reliability and practicality. While TFLN waveguides provide much tighter optical confinement and in principle stronger nonlinear interactions[43], current exploration of THz generation in the TFLN platform has been limited to pulsed operation with relatively low efficiency[44]. In short, there is still an unmet demand for systems capable of generating continuous THz waves in a compact, tunable fashion at room temperature.

In this study, we address the above challenges by demonstrating a monolithic integrated TFLN platform simultaneously capable of efficient THz-optic modulation and continuous THz wave generation at frequencies up to 500 GHz. The strong bidirectional interaction results from an optimized THz-optic co-design on a TFLN-on-quartz platform, which simultaneously features velocity- and impedance-matching, large overlap between THz and optical waves, and low THz transmission loss. We first present a THz-optic modulator with a low RF half-wave voltage of 8 V at 500 GHz, as well as ultrahigh 3-dB and 6-dB bandwidths of 145 GHz and 310 GHz, respectively. We then demonstrate a continuous-wave THz generator with conversion efficiencies exceed $10^{-6}$/W across a wide frequency range from 220 to 500 GHz, using only milliwatt-level pump power. This performance surpasses all previous tunable optical THz wave generation devices by at least tenfold when normalized to continuous wave efficiencies[35,36]. We further take advantage of the coherent nature of optical THz generation to demonstrate high-speed electro-THz modulation at up to 35 GHz and with various modulation schemes by modulating RF signals onto one of the optical pump lasers. The successful integration of efficient THz-optic modulator and THz generator on the same chip-scale platform paves the way for compact and multifunctional THz photonic systems.

**Results**

**Working principle and device design**

Figure 1 presents a schematic representation of the dual-functional monolithic integrated LN photonic platform, which supports strong and velocity-matched interaction between optical and THz waves. We use a metallic coplanar waveguide transmission line with a slow-wave electrode design, as depicted in Fig. 1(a). Although this choice involves higher THz transmission loss compared to dielectric waveguides, it allows deep sub-wavelength THz field confinement within the metallic electrode gap and significantly increases the nonlinear overlap between THz and optical modes. Figures 1(b) and 1(c) depict the simulated THz (500 GHz) and optical (200 THz) mode profiles ($E_z$), featuring strong field confinement at both frequencies and polarization both along the *z*-crystal direction, which allows the exploitation of the largest $\chi^{(2)}$ tensor component ($r_{33}$ and $d_{33}$) in LN. During the THz-optic modulation process, incident transverse-electric (TE$_0$) light from the left side is split into two branches of a Mach-Zehnder interferometer (MZI) and co-propagates with a THz wave confined in a ground-signal-ground (GSG) transmission line. Phase modulation accumulates along the two arms in a push-pull configuration, driven by the Pockels effect, and is finally converted into amplitude modulation when the two arms recombine. Figure 1(d) shows the frequency-domain schematic of the modulation process, which produces first-order sidebands separated from the carrier wave by the THz signal frequency. On the other hand, THz generation makes use of a similar device configuration but without an MZI structure, where two input light sources of closely spaced frequencies ($\omega_1$ and $\omega_2$) and identical polarization (TE$_0$) are combined and introduced into the optical waveguide from the same port. The strong nonlinear overlap between the two optical waves and the THz wave leads to frequency mixing and generation of the THz signal at the frequency separation $\omega_1 - \omega_2$, as illustrated in Fig. 1(e). The generated THz wave collinearly propagates and continuously accumulates inside the transmission line before finally being extracted from the output end by a GSG THz probe. In both processes, the rib LN waveguide (1.2 µm wide and 250 nm tall) facilitates effective optical confinement with low loss of below 0.1 dB/cm. We use an optimized electrode gap of 5.5 μm to balance optical loss and electro-optic interaction strength.

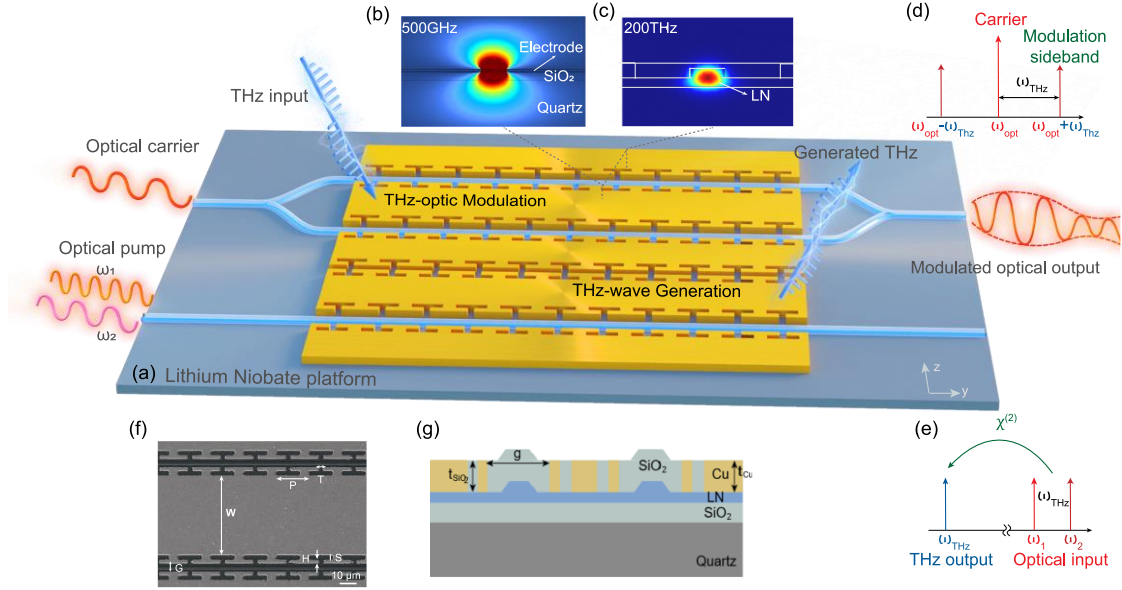

Fig. 1: Schematic of the dual-functional TFLN chip for THz-optic modulation and THz wave generation. (a) Schematic illustration of the device with strong optical confinement in the waveguides (blue) and THz confinement in the coplanar transmission lines (yellow), fostering a strong and velocity-matched interaction between the two modes. In THz-optic modulation, THz signals are fed into the transmission line via a probe and modulated onto the optical carrier. In THz-wave generation, two laser signals are input into the optical waveguide, resulting in frequency mixing and generation of THz wave, which is extracted via another probe. (b-c) Numerically simulated electric field distributions of the THz mode at 500 GHz (b) and the optical mode at 200 THz (c). (d-e) Frequency-domain illustrations of the THz-optic modulation (d) and THz wave generation (e) processes. (f) Scanning electron microscope image of the fabricated device. (g) Cross-sectional schematic of the device. The corresponding design parameters are shown in SI Note 1.

The key to an efficient and continuous build-up of THz-optic interaction in these devices is to achieve low THz propagation loss while maintaining velocity- and impedance-matching conditions. In our previous modulation TFLN-on-Si devices, the coplanar transmission lines see substantial propagation loss at THz frequencies (2 dB/mm at 500 GHz) due to significantly increased conductor (ohmic) loss in the narrow signal line (20 μm) and field leakage into the Si substrate[28]. In contrast, our design incorporates a 50 μm signal width and utilizes a quartz substrate to reduce the conductor loss and substrate leakage, respectively. To compensate for the rapid velocity of THz waves on the quartz substrate, we implement a slow-wave structure with periodical capacitively loaded electrodes to achieve matched THz phase velocity with optical group velocity. This ensures a coherent energy transfer between the optical and THz waves along the entire device length. By carefully engineering the T-shaped capacitive loading structures, we also achieve a near 50 Ω transmission line impedance to align with our external THz driving and detection circuits. Compared with previous reports[25,26], Our design utilizes a much shorter capacitive loading period of 20 μm such that each loaded element still effectively operates as a lumped capacitor even at an extremely high frequency of 500 GHz (wavelength = 600 μm). The more detailed engineering parameters are listed in SI Table 1. Figures. 1(f) and 1(g) show the top-view scanning-electron microscope (SEM) image and cross-sectional schematic of our fabricated device, respectively.

We first perform a detailed characterization of the basic electro-optic and electrical properties of our devices. The 8 mm long MZI electro-optic modulator shows a measured low-frequency $V_\pi$ of 3.04 V under triangular voltage sweeps, corresponding to a voltage-length product $V_\pi L$ of 2.43 V·cm, as shown in Fig. 2(a). This value is consistent with previous results in TFLN platforms and represents a good electro-optic overlap in our device[13,25]. We then evaluate the electrical performances, i.e., effective refractive index and propagation loss, of the fabricated slow-wave transmission line at THz frequencies using a vector network analyzer (VNA) and a frequency-extension module. Figure 2(b) displays the simulated and measured THz dispersion curves from 220 to 320 GHz, showing a THz phase index of ~ 2.26 (blue dots) matched with the optical group index (blue dashed line) for $H = 2$ μm. Our simulation results using the finite element method (FEM, Ansys HFSS) suggest a larger $H$ value corresponds to a higher RF effective index and a slower velocity of THz wave [solid lines in Fig. 2(b)], consistent with our experimental results.

From the measured scattering parameters (S-parameters), we also extract the electrical loss coefficient ($\alpha_{RF}$) of the transmission line at THz frequencies, as displayed in Fig. 2(c). The electrical loss stems from two primary sources: conductor loss ($\alpha_c$) and dielectric loss ($\alpha_d$). The conductor loss, $\alpha_c$, typically scales with the square root of the frequency, and the dielectric loss, $\alpha_d$, increases linearly with frequency. The actual measurement results are represented by black dots, while the blue dashed line corresponds to fitted electrical loss following:

$$S_{21} = (\alpha_c \sqrt{f} + \alpha_d f)L + A, \tag{1}$$

where $L$ is the electrode length, and intercept $A$ represents the low-frequency loss due to residual impedance mismatch. Our achieved low-frequency loss of 0.4 dB is significantly lower than the 2.6 dB value reported in our previous work thanks to the capacitively loaded electrode design[28]. This translates into a characteristic impedance ($Z_c$) of the transmission line is approximately 46 Ω, closely matching with the target 50 Ω impedance. Based on the fitting results, we estimate $\alpha_c = 0.264$ dB·cm$^{-1}$·GHz$^{-1/2}$ and $\alpha_d = 0.028$ dB·cm$^{-1}$·GHz$^{-1}$, both of which show good agreement with the previous work on quartz substrate[25]. Our TFLN-on-quartz device achieves a significantly lower overall THz propagation loss (14 dB/cm at 300 GHz) compared to our previous design on silicon (22 dB/cm) thanks to the wider signal line and the reduced substrate leakage[28]. Our measurement extrapolates to a propagation loss of approximately 2 dB/mm at 500 GHz, which is sufficient to support a total device length of 5-10 mm without excessive THz wave attenuation. The relatively low loss levels are crucial for facilitating a coherent accumulation of nonlinear interactions in both THz-optic modulation and THz wave generation processes.

According to these results, our design effectively meets all requirements for effective THz-optic interactions, including optimal nonlinear overlap, precise velocity matching between THz and optical waves, minimal THz propagation loss, and impedance matching. Next, we make use of these devices to demonstrate efficient THz-optic modulation, optical THz generation, and electro-THz modulation functions.

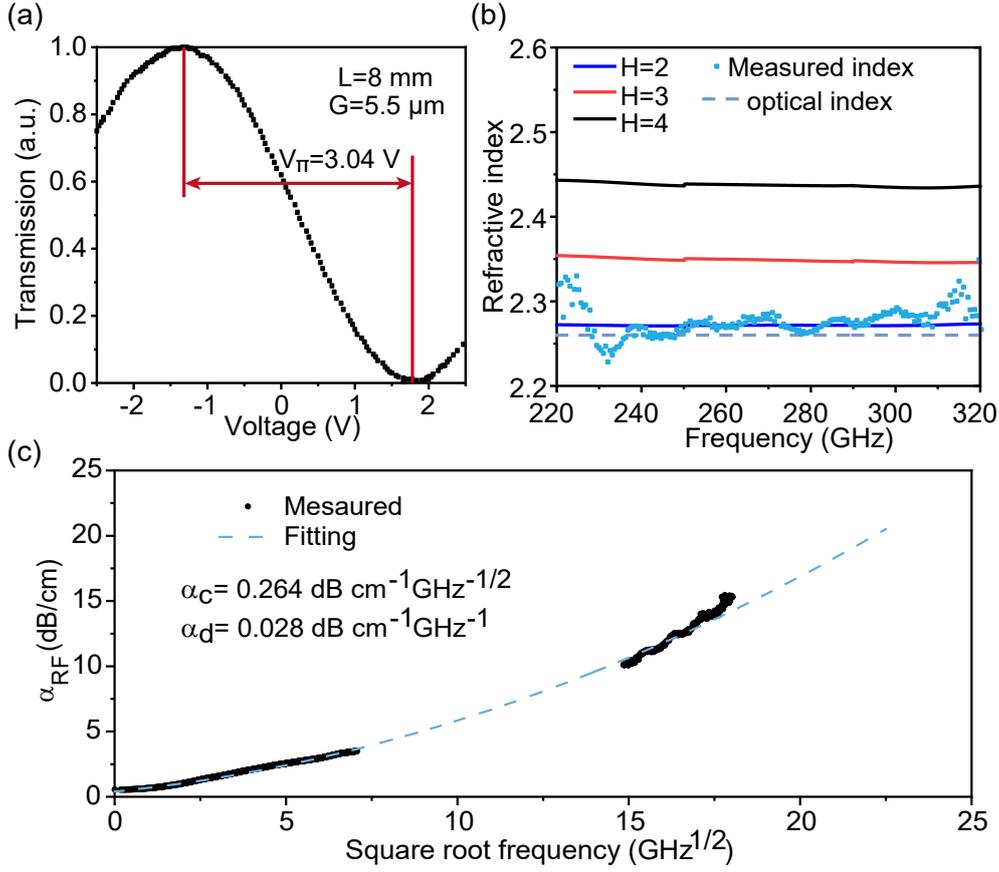

Fig. 2: Basic electro-optic and electrical characterizations. (a) Normalized optical transmission of the fabricated modulator as a function of applied low-frequency voltage. (b) Simulated THz effective (phase) indices for various $H$ values (solid lines) and the measured values (blue dots) across a frequency range from 220 GHz to 320 GHz. Blue dashed line shows the simulated optical group index. (c) Measured RF loss (black dots) and fitting results (blue dashed line) versus square-root frequency.

**On-chip THz-optic modulation**

We demonstrate low-voltage THz-optic modulation at frequencies up to 500 GHz using the testing setups shown in Figs. 3(a) and 3(b) following our previously established methodologies[28]. In an ideal scenario, increasing the length of the modulator always lowers the modulation voltage, which however comes at the cost of reduced tolerance to velocity mismatch, particularly at high frequencies. Here, we use an 8 mm long device to balance this trade-off. At high frequencies (10-500 GHz), the RF $V_\pi$ values are determined by analyzing the power ratios between the first-order sidebands and the carrier across different frequencies captured by an optical spectrum analyzer (OSA), as shown in Fig. 3(a). The raw measured data (blue circles) and the smoothed data for clearer visualization (purple line) in Fig. 3(c) show good alignment with the theoretical predictions (red dashed line) derived from the measured electrical S-parameters. The electro-optic $S_{21}$ parameters are obtained by comparing the measured RF $V_\pi$ values with DC $V_\pi$ following our theoretical model[28], which aligns well with direct $S_{21}$ measurements from a VNA at the lower-frequency band [10 MHz - 67 GHz, orange triangles in Fig. 3(d)]. This agreement validates the reliability and robustness of our measurement approach at THz frequencies.

Notably, our THz-optic modulator maintains low RF $V_\pi$ throughout the entire measured frequency range, i.e., 6 V at 300 GHz and 8 V at 500 GHz. In comparison, our previous TFLN device on a silicon substrate with a longer length of 10.8 mm exhibited an extrapolated RF $V_\pi$ of 28 V at 500 GHz[28]. This 3.5 times reduction in RF $V_\pi$ translates to a significant decrease in modulation power consumption by more than an order of magnitude, which is critical for practical application scenarios. Furthermore, our device achieves ultrahigh 3-dB and 6-dB bandwidths of 145 GHz and 310 GHz, respectively, significantly surpassing all measured bandwidths in previous LN-based modulators (see SI Table 2 for a detailed comparison). Combining the ultrabroad electro-optic bandwidths and low driving voltage, our THz-optic modulators could become a highly promising solution for 6G communications and advanced THz photonics applications.

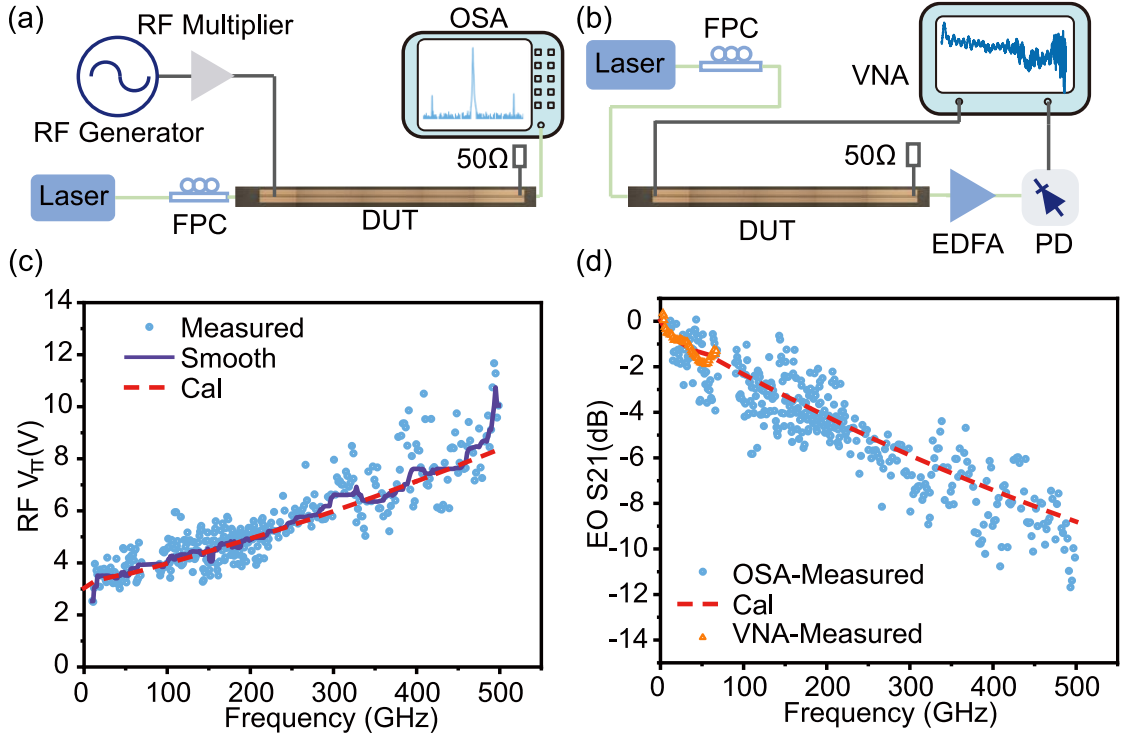

Fig. 3: THz-optic modulation performance. (a-b) Setups for measuring the electro-optic responses from 10 GHz to 500 GHz using an OSA (a) and from 10 MHz to 67 GHz by VNA (b). Insets show a microscope image of the full device. FPC, fiber polarization controller; EDFA, erbium-doped fiber amplifier; PD, photodetector; VNA, vector network analyzer; OSA, optical spectrum analyzer. (c-d) Measured and calculated RF $V_\pi$ (c) and electro-optic $S_{21}$ (d) of the fabricated modulator. Blue circles and orange triangles correspond to the raw measured data directly extracted from OSA and VNA measurements, respectively. Purple solid line shows smoothed measurement values for better comparison with the theoretical values (red dashed) calculated based on the measured electrical $S$-parameters.

**On-chip continuous THz wave generation**

We then demonstrate efficient, continuous-wave, and highly tunable THz generation across a broad frequency range from 220 GHz to 500 GHz (Fig. 4). Here a straight optical waveguide runs through one dielectric gap of the GSG transmission line instead of the MZI structure used earlier. This is mainly because the anti-symmetric THz mode profile would result in zero overall nonlinear overlap if the optical signals running in the two gaps have the same phase. Although this could be solved by adding a long optical delay line in one arm to introduce a $\pi$ phase in the nonlinear polarization $P^{NL} \sim E(\omega_1) E(\omega_2)^*$, the splitting of input optical power would lead to reduced optical intensity in each arm and ultimately the same overall conversion efficiency as our straight-waveguide design. Figure 4(a) shows the continuous THz wave generation setup. Figure 4(b) shows the measured THz signal spectra extracted at the output end of the transmission line at a wide range of frequencies, which can be simply controlled by adjusting the frequency difference between the two laser sources. The generated continuous THz wave power across the measured frequency range is consistently at the level of -65 dBm at a total on-chip optical pump power of 8 dBm, as illustrated in Fig. 4(c). Our THz generator does not see a significant efficiency drop even at the 500 GHz upper measurement limit imposed by our current probes and characterization equipment. This suggests the potential for further extending its operational frequency range beyond this limit. By placing the output probe at different locations of the transmission line, we can measure the generated THz power for different device lengths, i.e. 5 mm, 8 mm, and 10 mm (neglecting the effect from the remaining unused section). Figure 4(d) shows the simulated and measured conversion efficiencies for different lengths and at various frequencies. The highest measured THz generation efficiencies are $2.90\times10^{-6}$/W at 300 GHz (black squares), $4.13\times10^{-6}$/W at 400 GHz (red circles), and $4.86\times10^{-6}$/W at 500 GHz (blue triangles) for a device length of 10 mm. Theoretical analysis of the simulated conversion efficiency could be found in our earlier work[43] and in SI Note 3. Briefly, the normalized THz generation efficiency $\Gamma$ in a lossless system can be defined and calculated using the following equation:

$$\Gamma = \frac{P_{\text{THz}}}{P_{\text{opt1}} \cdot P_{\text{opt2}}} = \frac{2\omega_{\text{THz}}^2 d_{\text{eff}}^2}{n_{\text{opt}}^2 n_{\text{THz}} \varepsilon_0 c^3 A} \cdot L^2 \cdot \text{sinc}^2\left(\frac{\Delta k L}{2}\right), \qquad (2)$$

where $P_{\text{opt1,2}}$ are the optical powers of input lasers at $\omega_1$ and $\omega_2$, $P_{\text{THz}}$ is the output THz power at $\omega_{\text{THz}}$, $\Delta k$ is the wavevector mismatch between optical and THz waves, $L$ is the waveguide length, $\varepsilon_0$ and $c$ are the permittivity and speed of light in vacuum, $n_{\text{opt}}$ and $n_{\text{THz}}$ are the effective refractive indices of the optical and THz modes, $A$ is the effective spot area dictated by the nonlinear overlap between the three waves:

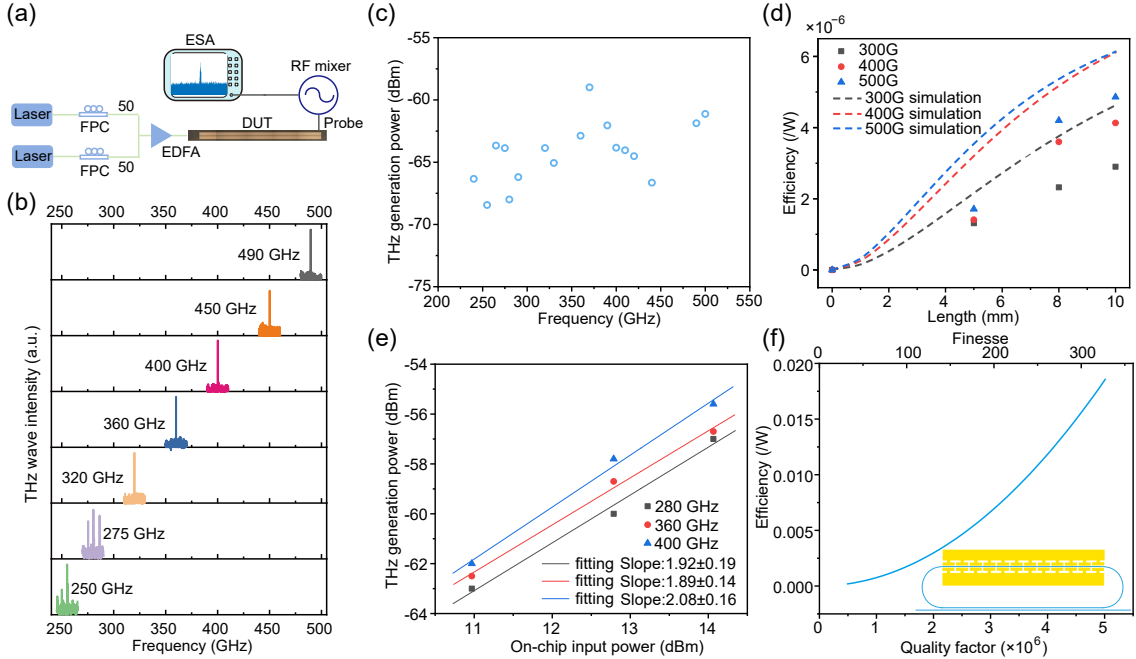

Fig. 4: THz wave generation performance. (a) Setup for THz wave generation measurement. Inset shows a microscope image of the full device. FPC, fiber polarization controller; EDFA, erbium-doped fiber amplifier; ESA, electrical spectrum analyzer. (b) Measured THz wave spectra at different frequencies ranging from 250 to 490 GHz. (c) Generated THz wave power at various frequencies for a fixed on-chip pump power of 8 dBm. (d) Conversion efficiency versus transmission line length at various frequencies. Black squares, red circles, and blue triangles represent 300, 400, and 500 GHz measurements. Dashed curves show the simulated results at corresponding frequencies. (e) THz wave generation power as a function of the input light power at different

frequencies in double logarithmic scales, together with linear fitting lines with slopes ~ 2 verifying the quadratic power dependence. (f) Simulated conversion efficiency versus quality factor (Finesse) at 500 GHz in a resonant structure. The inset shows the device schematic.

$$A = \frac{A_{opt}^2 \cdot A_{THz}}{A_{overlap}^2} = \frac{\left(\int_{all}|\mathbf{E}_{opt}|^2 dxdy\right)^2 \left(\int_{all}|\mathbf{E}_{THz}|^2 dxdy\right)}{\left|\int_{LN}|E_{opt,z}|^2 E_{THz,z}^* dxdy\right|^2}, \qquad (3)$$

where $\int_{LN}$ and $\int_{all}$ denote cross-section integration over the LN region only and all space, respectively. $\mathbf{E}_{opt}$ and $\mathbf{E}_{THz}$ are the electric fields of the optical and THz modes and $E_{opt,z}$ and $E_{THz,z}$ are their corresponding z components that utilize the largest nonlinear coefficient $d_{33}$. As Eq. (2) shows, the THz generation efficiency inherently increases quadratically with device length when the THz transmission loss is not significant. Remarkably, our device shows a strong upward trend in THz conversion efficiency even at a 10-mm length, which also matches well with our full theoretical model [dashed lines in Fig. 4(d)] that takes into account the measured THz loss [Fig. 2(c)]. This trend suggests that there is potential for further efficiency gain with the use of longer electrodes[45]. The small discrepancies between simulated and experimental efficiencies may be attributed to larger-than-expected on-chip and facet optical losses and fabrication-induced deviation of structural parameters (e.g., metal gap). Despite this, our measured efficiencies are more than ten times higher than those of current LN-based tunable THz generators when normalized to continuous-wave values[35,36] (see SI Table 3 for detailed comparison). We further evaluate the input-output power relationship by concurrently increasing the input power of the two lasers from 12.5 mW to 25.5 mW (11 dBm to 14 dBm) at THz frequencies of 280, 360, and 400 GHz, as shown in Fig. 4(e). Here, the power levels correspond to total on-chip optical powers and the two lasers are kept at 1:1 power ratio. The output THz powers exhibit good quadratic relationships with input optical powers among these three frequencies, as evidenced by the linear fitting curves in the double logarithmic plot in Fig. 4(e) with slopes of approximately 2. Currently, the generated THz signals see relatively large linewidths since the two pump lasers are not frequency locked, as detailed in SI Note 4, which could be substantially improved by locking the two lasers via electro-optic or Kerr frequency combs[46,47]. Compared to other technologies requiring up to dozens of watts of pump power[34-36], our system could achieve microwatt-level outputs using sub-watt input power, significantly reducing power consumption while maintaining excellent frequency tunability.

The THz wave generation efficiency could, in principle, be further enhanced by embedding the system in a racetrack optical resonator[43], as schematically shown in the inset of Fig. 4(f). Thanks to the non-resonant nature of the THz mode, no additional phase and velocity matching conditions are required as long as the optical pump frequencies are matched to the cavity resonances. This resonator configuration dramatically increases effective pump powers by the cavity finesse. Here, we assume that a racetrack resonator with a bending radius of 80 μm, a straight section length of 5 mm, and a free-spectral range (FSR) of ~12 GHz is critically coupled with loaded $Q$ factors ranging from $0.5 \times 10^6$ to $5 \times 10^6$, which have well been achieved in many recent reports in the TFLN platform[48,49]. The effective pump power experiences an amplification from 10 to 104 times inside the resonator compared to the power in the bus waveguide, leading to an overall conversion efficiency enhancement of over 10,000 times at a $Q$ factor of $5 \times 10^6$. This corresponds to a remarkable theoretical THz generation efficiency of 1.85%/W at 500 GHz. Although this resonant system allows THz emission only at discrete frequencies (with spacing of FSR ~ 12 GHz), the ability to generate continuous-wave THz signals at a wide range of stepped frequencies (e.g., 200 – 500 GHz) could still offer significant potential for many applications including communications and imaging.

**High-speed electro-THz modulation**

Finally, we showcase the versatility of our optical THz wave generation scheme by demonstrating high-speed modulation of the generated THz waves via mature and cost-effective optical modulation techniques. Figure 5(a) details our experimental setup for electro-THz modulation, where one of the optical pump signals is first modulated by a standard commercial electro-optic modulator before mixing with the other pump laser with a center frequency difference of 420 GHz. Figure 5(b-d) illustrates the measured electrical spectra of the modulated THz wave at various frequencies (up to 35 GHz) and modulation states (carrier maintained/suppressed modulation), whereas Fig. 5(e-g) show the corresponding modulated input optical spectra. The power ratios between the carrier and sidebands are highly consistent between the optical and electrical spectra, e.g., 7.96 dB optically and 8.08 dB electrically at 35 GHz [Figs. 5(c) and 5(f)]. The high extinction ratio in the carrier-suppressed modulation signal when biasing the modulator at null [Fig. 5(d)] also confirms the excellent coherence and linearity of the optical-THz conversion process. Notably, achieving THz wave modulation at such frequencies is typically non-trivial and necessitates costly THz mixers. Our approach utilizes established and readily available optical methods, currently limited only by the 40 GHz bandwidth of the commercial electro-optic modulator used. The electro-optic modulation block could be further integrated on the same TFLN chip, leveraging the excellent scalability of our platform, leading to fully integrated THz generation and modulation systems with even higher modulation speed, smaller footprint and lower cost.

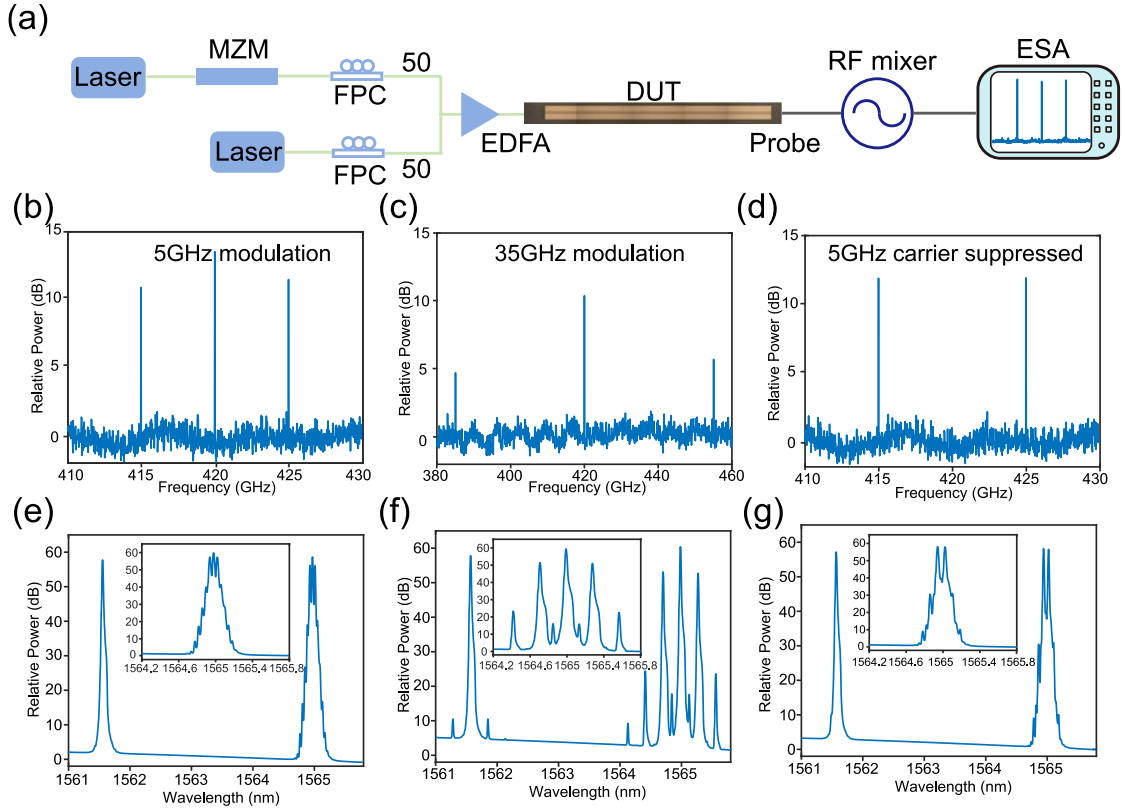

Fig. 5: High-speed electro-THz modulation. (a) Schematic diagram of the experimental setup. MZM, Mach-Zehnder modulator. (b-d) Measured electrical spectra of the modulated THz waves for various modulation frequencies and formats. (e-g) Corresponding optical spectra of the modulated pump signal. (b) and (e) correspond to an amplitude modulation frequency of 5 GHz, (c) and (f) correspond to a modulation frequency of 35 GHz, (d) and (g) show the spectra of 5 GHz carrier-suppressed modulation.

## Discussion

In summary, we have demonstrated a dual-functional TFLN photonic integrated circuit that can simultaneously perform THz-optic modulation and continuous THz wave generation. The strong bidirectional THz-optic interaction in our platform allows to achieve efficient THz-optic modulation with low RF $V_\pi$ values of 6 V at 300 GHz and 8 V at 500 GHz, as well as record-high continuous THz wave generation efficiency surpassing $10^{-6}$/W across a broad frequency range from 220 to 500 GHz. The excellent coherence in the THz-optic conversion processes provides a highly effective and versatile toolbox for the generation, transmission, and processing of THz signals in the optical domain, namely THz photonics, potentially benefiting a variety of applications that require advanced signal encoding and decoding. Here, we provide the first proof of this concept by realizing the high-speed modulation of THz signals via an optical approach. Further integration of functional optical and THz elements on the same platform, including amplitude and phase modulators, optical filters, delay lines, and on-chip antennas, could enable highly integrated and multi-functional THz photonic systems with applications in high-speed THz communications and high-resolution THz ranging and imaging.

## Methods

**Device design and simulation.** The propagation properties of the THz wave along the transmission line were analyzed using the finite element method (FEM) in Ansys HFSS. We simulated performance parameters such as the THz wave's refractive index and the characteristic impedance over an electrode length of 1000 μm, then scaled these parameters up to match the actual required electrode length. Detailed parameters of our device design can be found in SI Note 1. The THz mode profile in Fig. 1(b) was computed through an FEM in Comsol Multiphysics, whereas the optical mode profile in Fig. 1(c) was generated using a commercial Finite Difference Eigenmode (FDE) solver (Lumerical, Mode Solutions). Our design incorporated a slow wave structure within the transmission line to achieve velocity matching between the THz and optical waves, while targeting a characteristic impedance close to 50 Ω. We adopt the same structure and principles as in the previous paper[25], while using a smaller capacitive loading period to increase the cut-off frequency. We numerically simulated the slow wave effect by adjusting the height (H) within the metal electrode slot using the finite element method (FEM, Ansys HFSS) and extracted the THz effective index from the numerically converged propagation constant[28]. We evaluated the THz generation conversion efficiency by numerically calculating the overlap between the simulated THz and optical mode profiles. This calculation considered both the effects of velocity matching and the experimentally measured THz wave propagation losses within actual devices.

**Device fabrication.** Devices were fabricated using a commercial x-cut LN wafer from NANOLN. This wafer consists of a 500-nm-thick LN, a 2-μm-thick buried oxide (BOX) layer, and a 500-μm-thick quartz substrate. Waveguide patterns were initially defined using electron-beam lithography (EBL) on hydrogen silsequioxane (HSQ) resist. The patterns were then transferred into the TFLN by dry etching 250 nm of the LN film using an Argon-plasma ($Ar^+$) based reactive ion etching (RIE) process. The fabricated rib waveguides have a height and top width of 250 nm and 1.2 μm, respectively. Subsequently, a layer of $SiO_2$ was deposited over the LN chip as a cladding layer using plasma-enhanced chemical vapor deposition (PECVD). Slow-wave electrodes were constructed through a series of processes, including photolithography, RIE, thermal evaporation, and a lift-off process. The final steps involved dicing and polishing the chip edges to enhance fiber-chip coupling.

**Characterization of THz wave transmission line.** The *S*-parameters of the slow wave transmission lines were measured using VNA with frequency extension modules. The measurements spanned from 10 MHz to 50 GHz using a VNA (Keysight E5080B), and from 140 to 220 GHz using another VNA (Agilent N5230C) with millimeter wave VNA Extender (V05VNA2-T/R). GSG probes for respective frequency bands were used for THz signal input and output. To ensure the accuracy of these measurements, the probes were calibrated to the tips using a commercial Short-Open-Load-Through (SOLT) calibration kit, which effectively removed the influence of any unwanted parasitic effects.

**Characterization of THz-optic modulation.** The electro-optic characterization was performed in the telecom C-band using a tunable-wavelength laser source (Santec TSL-510). A three-paddle polarization controller was used to ensure TE mode excitation. Light was coupled into and out from the chip under test using tapered lensed fibers, with a coupling loss of ~ 5 dB/facet. For the measurement of DC/low-frequency $V_\pi$, a kilohertz electrical sawtooth waveform was generated from an arbitrary-waveform generator (AWG, RIGOL DG4102) and applied to the transmission line of the electro-optic modulator through a GSG probe (GGB industries, 50 GHz). The output optical signal of the electro-optic modulator was detected using a photodetector (New Focus 1811) and recorded on an oscilloscope (RIGOL DS6104). For high-speed THz-optic modulation performance of RF $V_\pi$, measurements were separately conducted in five frequency bands ranging from 9 to 500 GHz using different signal generation methods. For the lowest band (<67 GHz), modulation electrical signals were directly generated from an RF generator (MG3697C, Anritsu). For higher frequencies (>67 GHz), RF signals were up-converted and amplified by frequency multipliers to the respective bands, i.e., 90-140 GHz, 140-220 GHz, 220-320 GHz, 320-500 GHz. The modulator was biased at the quadrature point, generating an output optical signal with two sidebands separated from the carrier by the THz frequency. The RF $V_\pi$ was tested by monitoring the power ratio between the sideband and the carrier, using an OSA (Yokogawa AQ6370)[28]. For direct electro-optic response measurement between 1-50 GHz, a small RF signal from VNA (Keysight, E5080B) was input into the modulator electrodes through a GSG probe. The output signals at various frequencies were captured by a high-speed photodetector (50 GHz photodetector, XPDV21x0) and sent to the other port of the VNA. Probe loss, RF cable loss and PD response were extracted from the measured electro-optic response.

**Characterization of THz wave generation.** THz wave generation process was characterized by using two continuous wave laser sources with certain frequency differences. The tunable telecom lasers (Santec TSL-510 and Santec TSL-710) were individually polarized into the $TE_0$ mode using three-paddle fiber polarization controllers. The signals were then combined through a 50/50 coupler, amplified using an erbium-doped fiber amplifier (EDFA, Amonics, AEDFA-L-30-B-FA), which also served as a regulator of the on-chip optical power. Tapered lensed fibers were used to couple light into and out from the waveguide facets of the LN chip. The generated THz signals were collected by a high-speed GSG probe and underwent frequency down-conversion via spectrum analyzer extension modules (SAX, VDI, WR3.4SAX-M & WR2.2SAX-M). The down-converted signal was then sent to the local oscillator of an Electromagnetic Spectrum Analyzer (ESA) for precise measurement of the THz signal frequency and power. We calibrated and deducted losses incurred from various components of the entire testing system, including the probe, waveguide, cable and SAX. This ensured the accuracy and reliability of our measurement.

For the measurement of the electro-THz modulation process, we utilized a commercial electro-optic modulator (EOSPACE, AX-0MVS-40) with RF and DC inputs to modulate one input laser. We applied different RF input power, frequency and DC voltage to adjust the modulation state. The continuous wave THz signal generated through optical methods can transfer the modulation information carried by the optical signal to the THz frequency band. The modulated THz signal was down converted using the SAX and monitored using an ESA.

**Funding.** Research Grants Council, University Grants Committee (CityU 11204820, CityU 11212721; CityU 11204022); Croucher Foundation (9509005); Hong Kong Institute for Advanced Study (HKIAS).

**Acknowledgments.** We thank Dr. Wing-Han Wong at CityU for their help in device fabrication.

**Conflict of interest.** The authors declare no conflicts of interest.

**Data availability.** Data underlying the results presented in this paper are not publicly available at this time but may be obtained from the authors upon reasonable request.

# Supplementary Information
# Monolithic lithium niobate photonic chip for efficient terahertz-optic modulation and terahertz generation

**Supplementary Note 1: Detailed parameters of devices**

The fabrication process of our samples is detailed in the methods section. The dimensions of the device are listed in the SI Table. 1.

SI Table 1: List of device parameters

| Thickness of quartz substrate | $t_{sub}$ | 500 μm |
|---|---|---|
| Thickness of bottom silicon oxide layer | $t_{SiO2,b}$ | 2 μm |
| Thickness of lithium niobate layer | $t_{LN}$ | 500 nm |
| Thickness of top silicon dioxide layer | $t_{SiO2,t}$ | 500 nm |
| Thickness of lithium niobate ridge waveguide | $t_{LN,w}$ | 250 nm |
| Width of lithium niobate ridge waveguide | $w_{LN}$ | 1.2 μm |
| Gold thickness | $t_{Cu}$ | 500 nm |
| Signal width | $w$ | 50 μm |
| Electrode gap | $G$ | 5.5 μm |
| Slot period | $P$ | 20 μm |
| Period gap | $T$ | 2 μm |
| Slot height 1 | $H$ | 2 μm |
| Slot height 2 | $S$ | 2 μm |

**Supplementary Note 2: Comparison with other thin film lithium niobate (TFLN) modulators**

SI Table. 2: List of TFLN modulators comparison.

| Ref | Platform | THz RF $V_\pi$ (V) | BW 3dB (GHz) | BW 6dB (GHz) | $V_\pi L$ (V·cm) | Modulation length (mm) |
|---|---|---|---|---|---|---|
| Our work | TFLN/Quartz | 6@300GHz<br>8@500GHz | 145 | 310 | 2.43 | 8 |
| 1 | TFLN/Silicon | N/A | 45 | N/A | 2.20 | 33 |
| 2 | TFLN/Silicon | 7.3@250GHz<br>28@500GHz* | 100 | 175 | 2.13 | 10.8 |
| 3 | TFLN/Silicon | N/A | 43 | N/A | 2.20 | 20 |
| 4 | TFLN/Silicon | N/A | 55 | N/A | 2.74 | 10 |
| 5 | TFLN/Silicon | N/A | > 67 | N/A | 2.25 | 22.5 |
| 6 | TFLN/Silicon | N/A | >60 (predicted) | N/A | 2.20 | 10 |
| 7 | TFLN/Silicon | N/A | 119 | N/A | 1.45 | 4 |
| 8 | TFLN/Silicon | 40@500GHz | N/A | N/A | 3.8 | 15.0 |
| 9 | TFLN/Quartz | N/A | 180 (extrapolate) | N/A | 2.70 | 20 |

BW: bandwidth

*: extrapolated

**Supplementary Note 3: Theoretical model of generated terahertz (THz) efficiency**

To achieve an efficient THz wave generation process with coherent signal buildup throughout the entire device length, it is important to satisfy the phase-matching condition. In our collinear guided-wave configuration, the condition is expressed as $\Delta k = k_{opt1} - k_{opt2} - k_{THz} = 0$, where $k_{opt1,2}$ and $k_{THz}$ are the wavevectors of the two pump modes and the THz mode, respectively.

Initially, we analyze the scenario of a lossless waveguide. In this context, the THz generation efficiency $\Gamma$ (normalized by pump optical power) could be calculated using the following equation:

$$\Gamma = \frac{P_{THz}}{P_{opt1} \cdot P_{opt2}} = \eta L^2 \cdot \mathrm{sinc}^2\left(\frac{\Delta k L}{2}\right), \tag{1}$$

where $P_{opt1,2}$ are the optical powers of input lasers at $\omega_1$ and $\omega_2$, $P_{THz}$ is the output THz power at $\omega_{THz}$, $L$ is the waveguide length. The normalized conversion efficiency, $\eta$, which is normalized by both pump power and device length, is determined by the following expression:

$$\eta = \frac{2\omega_{THz}^2 d_{eff}^2}{n_{opt}^2 n_{THz} \varepsilon_0 c^3 A}, \tag{2}$$

where $\varepsilon_0$ and $c$ are the permittivity and speed of light in vacuum, respectively, $\omega_{THz}$ is the angular frequency of the intended THz wave, $n_{opt}$ and $n_{THz}$ are the effective refractive indices of the optical (assuming $n_{opt1} \approx n_{opt2}$) and terahertz fundamental TE modes, respectively, and $d_{eff} = d_{33} = \chi^{(2)}/2 = 195$ pm/V is the effective nonlinear susceptibility at THz range, $A$ is the effective spot area between the three waves:

$$A = \frac{A_{opt}^2 \cdot A_{THz}}{A_{overlap}^2} = \frac{\left(\int_{all} |\mathbf{E}_{opt}|^2 dxdy\right)^2 \left(\int_{all} |\mathbf{E}_{THz}|^2 dxdy\right)}{\left|\int_{LN} |E_{opt,z}|^2 E_{THz,z}^* dxdy\right|^2}, \tag{3}$$

where $\int_{LN}$ and $\int_{all}$ denote cross-section integration over the LN region only and all space, respectively. $\mathbf{E}_{opt}$ and $\mathbf{E}_{THz}$ are the electric fields of the optical and THz modes and $E_{opt,z}$ and $E_{THz,z}$ are their corresponding $z$ components that utilize the largest nonlinear coefficient $d_{33}$.

When taking into account the propagation loss of the THz mode, the conversion efficiency expression in Eq. (2) can be modified by adding an imaginary part to the phase mismatch as $\Delta k' = \Delta k - i\alpha/2$, where $\alpha$ is the THz attenuation coefficient. The conversion efficiency in lossy medium is then given by:

$$\Gamma' = \eta L^2 \cdot \mathrm{sinc}^2\left(\frac{\Delta k' L}{2}\right) \cdot e^{-\alpha L} = \frac{2\omega_{THz}^2 d_{eff}^2}{n_{opt}^2 n_{THz} \varepsilon_0 c^3 A} \cdot \mathrm{sinc}^2\left(\frac{(\Delta k - i\alpha/2)L}{2}\right) \cdot L^2 \cdot e^{-\alpha L}, \tag{4}$$

As a result, to optimize the overall conversion efficiency, it is essential to minimize the effective spot area $A$ while maintaining a reasonable THz loss.

**Supplementary Note 4: Linewidth of the generated THz signal**

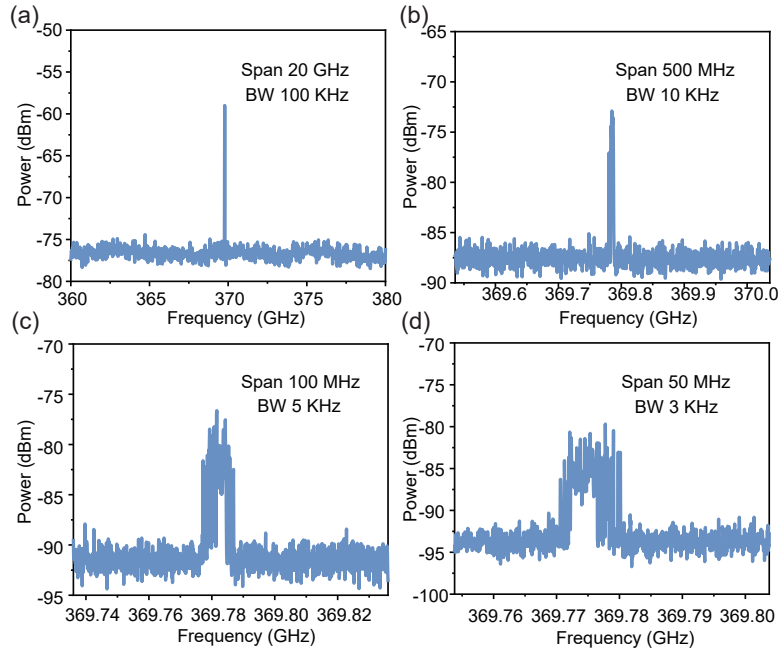

SI Fig. 1: Measured THz output spectra under different measurement span and bandwidth (BW) at 370 GHz. (a) Span 20 GHz, BW 100 kHz; (b) Span 500 MHz, BW 10 kHz; (c) Span 100 MHz, BW 5 kHz; (d) Span 50 MHz, BW 3 kHz.

In our current implementation, the linewidth and noise of the generated THz signal is mainly limited by the frequency instability of the two unlocked pump lasers. Our Santec laser sources exhibit a wavelength instability of 0.1 pm (12 MHz) and an instantaneous linewidth narrower than 100 kHz. This laser instability is directly translated into the frequency instability of the generated THz signals, as we observe in SI Fig. 1 when reducing the measurement resolution bandwidth to a few kHz. The spectral envelope captured during low-speed scans with small resolution bandwidth [SI Fig. 1(d)] shows a frequency drift typically between 10-20 MHz, mirroring the inherent instability of the laser's output. Addressing this challenge could involve shifting to laser sources with greater precision and stability, or implementing external mechanisms to synchronize the two lasers.

**Supplementary Note 5: Comparison with other terahertz generation schemes**
We have conducted a comparison of LN-based THz generation reports, focusing on efficiency, tunability, and operational features, as detailed in SI Table. 3. Most pulsed THz generation schemes [Suppl. Refs. 12-15] feature rather low conversion efficiencies when normalizing the efficiency to a continuous wave case, typically on the order of or below $10^{-11}$/W. Among existing continuous wave THz generation techniques, the approach reported by Scheller et al. [Suppl. Ref. 9] boasts high efficiency. However, it is restricted to offering outputs at only two fixed frequencies (1T and 1.9T), lacking the capability for frequency tunability. While alternative continuous wave schemes with tunable devices (Kiessling et al. and Sowade et al. [Suppl. Refs. 10, 11]) achieve efficiencies slightly exceeding $10^{-7}$/W, they require pump powers in the tens of watts range for nonlinear generation, resulting in significant energy consumption. These limitations, coupled with their bulkiness, lack of portability, and complex experimental setups, hinder their practical implementation in THz applications.

In contrast, our on-chip approach achieves the highest tunable continuous THz wave generation efficiency ($4.8\times10^{-6}$/W). The advantage of our method in terms of efficiency and tunability, presenting a compelling solution for practical THz wave generation applications.

SI Table. 3: List of LN-based THz generation comparison.

| Ref | Method | CW | Tunable | Broadband/single frequency | Room temperature | Efficiency (normalized to CW scheme) (/W) |
|---|---|---|---|---|---|---|
| Our work | On-chip | ✓ | ✓ | single | ✓ | $4.8 \times 10^{-6}$ |
| 10 | bulk | ✓ | × | single | ✓ | $3.2 \times 10^{-6}$ |
| 11 | bulk | ✓ | ✓ | single | ✓ | $1.67 \times 10^{-7}$ |
| 12 | bulk | ✓ | ✓ | single | ✓ | $1.8 \times 10^{-7}$ |
| 13 | bulk | × | ✓ | single | × | $3.56 \times 10^{-13}$ |
| 14 | bulk | × | × | broadband | ✓ | $2.81 \times 10^{-11}$ |
| 15 | bulk | × | ✓ | single | ✓ | $6.25 \times 10^{-19}$ |
| 16 | bulk | × | × | single | × | $2.27 \times 10^{-11}$ |

CW: continuous wave